\documentclass{article}
\usepackage{geometry}
\usepackage[bitstream-charter]{mathdesign}
\usepackage{paralist,todonotes,microtype}
\usepackage[T1]{fontenc}
\usepackage[utf8]{inputenc}
\usepackage{listings}
\usepackage{float}
\usepackage{url}
\usepackage[colorlinks,allcolors=blue]{hyperref}

\title{Avast-CTU Public CAPE Dataset\footnote{Available at \url{https://github.com/avast/avast-ctu-cape-dataset}.}}
\author{Branislav Bo{\v s}ansk{\' y}$^{1,2}$, Dominik Kouba$^{2}$, Ond{\v r}ej Ma{\v n}hal$^{2}$, Thorsten Sick$^{1}$,\\Viliam Lis{\'y}$^{1,2}$, Jakub K{\v r}oustek$^{1}$, Petr Somol$^{1}$\\~\\
$^1$Avast Software\\
{\small \texttt{\{branislav.bosansky, thorsten.sick, viliam.lisy, jakub.kroustek, petr.somol\}}}\\
{\small \texttt{@avast.com}}\\~\\
$^2$Artificial Intelligence Center, Dept. of Computer Science, \\Faculty of Electrical Engineering, Czech Tecnical University in Prague\\
{\small \texttt{\{branislav.bosansky, dominik.kouba, ondrej.manhal, viliam.lisy\}@fel.cvut.cz}}\\
} 
\date{}

\begin{document}

\maketitle

\section{Introduction}
Millions of new malicious software emerge every day worldwide~\footnote{See, for example, statistics at \url{https://portal.av-atlas.org/malware/statistics}} threatening to compromise consumer devices, stealing private data, and/or cause financial losses~\footnote{\url{https://cybersecurityventures.com/hackerpocalypse-cybercrime-report-2016/}}.
These new threats put increasing pressure on malware-detection software to continuously improve detection capabilities in order to provide reliable protection to the users.

There is a plethora of methods for detecting malicious samples~(e.g., see surveys \cite{ML-survey,dynamic_survey}). 
Broadly speaking, we can distinguish two main categories: (1) detecting the samples based on their \emph{static features} and (2) detecting the samples based on a \emph{behavioral analysis}.
The static features typically consist of considering the whole sample (e.g., as an image~\cite{ml-image}) and/or properties of its most important parts (e.g., by examining in details header of a Windows portable executable (PE) file)~\cite{sikorski2012practical}.
The behavioral analysis consists of executing (or simulating the execution) of the sample and logging performed actions in order to determine whether these actions have characteristics of malicious behavior~\cite{dynamic_survey}. 
The main advantage of the first approach is the computational efficiency since extracting static features from the file itself can be much faster compared to the (simulated) execution.
On the other hand, the main disadvantage of the static approach is the inability to reliably distinguish malicious samples from benign samples in case the sample is encrypted and/or the clean file is altered in a minor way to exhibit malicious behavior. 
The methods relying on behavioral analysis can discover malicious behavior even in encrypted samples, however, they require significantly more resources to run or simulate the instructions of the analyzed sample.
In either case, the growing number of new, previously unseen samples makes the usage of automated decision/classification methods of artificial intelligence (AI) and machine learning (ML) inevitable in the malware-detection domain.

Framing the problem of malware detection as an AI/ML problem reveals interesting and unique properties of the domain that are less prevalent in other domains. 
These properties, among others, include:
\begin{itemize}
    \item the constantly changing distribution of evaluated data -- changes are induced by authors of new benign software as well as active attempts of malware authors to avoid the detection
    \item noisy and uncertain labels -- determining the ground-truth label can be extremely costly in some cases in time and the required level of expertise of the human analyst performing the evaluation; therefore, there can be considerable noise in labels in large-scale datasets~\cite{joyce21}
    \item focus on the performance of the classifiers under very low false-positive rates~\cite{Marculet19} -- since many malware-detection systems operate using a layered architecture~(e.g., see \cite{Saiyed18}), a false negative on one layer can be resolved by a different decision-making module / machine-learning classifier on a different layer. On the other hand, a false positive is reported immediately. Due to a large number of clean files, the false-positive rate of the classifier has to be as low as $\approx10^{-4}$.
\end{itemize}
The presence of such unique properties of the malware-detection domain from the AI/ML perspective requires the researchers to focus specifically on these domains.
To do so, it is important to have enough data that exhibit these unique features and allow researchers to develop and evaluate new methods that improve the detection capabilities of new models and advance the state of the art.

While several datasets for malware-detection problems were published over the past years~\cite{anderson2018ember,ronen2018microsoft,harang2020sorel20m}, the focus was primarily on static features of malicious samples.
Malware-detection datasets focusing on behavioral data are scarce. 
A few existing exceptions include a dataset of features extracted from API Calls~\cite{APIDS,Oliveira2019} or statistics based on the first few seconds of execution~\cite{RHODE2018578}.
However, the existing dynamic datasets offer only a limited viewpoint on the behavior of the samples. 
On the other hand, there are public sandbox projects~\cite{oktavianto2013cuckoo,capev2} that create a high-fidelity virtual environment for executing analyzed samples and as a result, they provide detailed analyses of the execution of the sample, including a complete tree of spawned processes, all API calls with all parameters, memory dump, etc.
Unfortunately, to the best of our knowledge, there is no such publicly available rich behavioral dataset of malicious samples. 
Therefore, our main contribution is collecting and releasing the first version of \textsc{Avast-CTU Public CAPE Dataset} consisting of JSON reports produced by CAPEv2 sandbox.

The released dataset contains $48,976$ malicious samples classified into 10 different malware families. The majority of the samples were collected mainly throughout the years from 2017 to 2019 and their counts vary in malware-family assignment and time. Such a dataset offers a representative subsample of actual malware data throughout a longer period of time. It can be used for classical supervised training of ML models but also for studying changes in data distribution over time (known as the concept drift), or trying to find the least costly parts of the reports to provide sufficiently accurate and robust classifier (e.g., using mainly static features to mimic the performance of the classifier that also uses behavioral features).

In this paper, we describe the dataset, provided metadata, and give a baseline classifier for classifying samples into malware families when data are split according to time. 

\section{CAPEv2 Sandbox}

CAPEv2 is an open-source sandbox~\cite{capev2} for execution analysis of samples that started as a fork of the original Cuckoo sandbox in 2019 and it is still actively developed and maintained at the time of collecting the reports. It further extends the monitoring and execution capabilities of the Cuckoo sandbox, including unpacking and/or capturing payload that provides more detailed information about executed samples and, for example, allows detection based on Yara signatures.

The actual execution of the samples and collection of the reports was performed at the Czech Technical University in Prague in July and August 2021, we used the recent version of CAPEv2 sandbox at that time. We ran the collection on 3 separate physical machines, each hosting 4 guest virtual machines that executed malware samples and were monitored by CAPE. These virtual machines were configured as follows: \begin{itemize}
    \item Windows 7 operating system was installed together with MS Office
    \item the virtual machines were altered to mimic a real personal computer\footnote{Setup was prepared such that it passed Pafish test, \url{https://github.com/a0rtega/pafish}.}, hence 
    \begin{itemize}
        \item a collection of typical popular applications was installed, including Google Chrome, Firefox, Adobe Reader, Spotify 
        \item private key was generated and stored in a standard location
        \item at least one password was stored in Chrome
        \item a collection of publicly available documents and images were downloaded and stored in typical locations
        \item external Internet connection was allowed
    \end{itemize} 
\end{itemize}

\begin{figure}[t]
    \centering
    \includegraphics[width=1.0\textwidth]{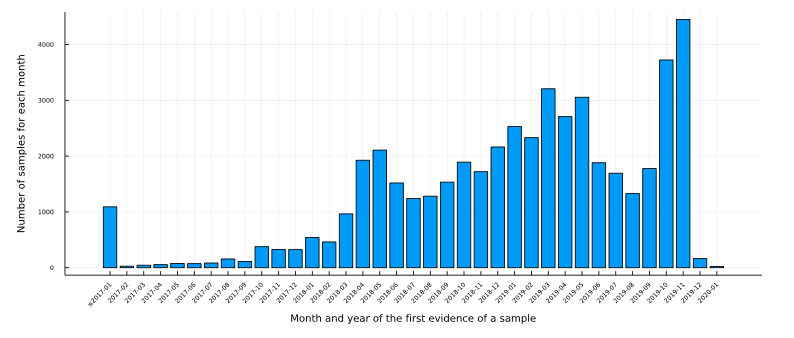}
    \caption{Histogram of months in which the samples in the dataset have been collected for the first time. The first month (2017-01) contains also all older samples.}
    \label{fig:times}
\end{figure}

\section{Data Description}
The purpose of the dataset is to allow researchers to work with detailed and rich behavioral data collected from several malware families over an extended period of time. We first describe the metadata followed by the structure of the samples themselves. 

\subsection{Metadata}
The dataset consists of $48,976$ samples. For each sample, we provide:
\begin{itemize}
    \item \textbf{sha256} of the sample for the identification,
    \item \textbf{classification} to a malware family,
    \item \textbf{type} of the malware sample,
    \item \textbf{date} of detection of the file.
\end{itemize}
All of them were classified as malicious, however, they belong to different malware families. Labeling these samples and assigning a particular malicious PE into a malware family is not an easy task as some code can be shared by various families thus causing possible noise in labels. According to the records in Avast systems, the classification of these samples into $10$ different malware families was as follows:
\begin{table}[h]
    \centering
    \begin{tabular}{|c|c|c|c|c|c|c|c|c|c|}
        \hline
        Adload & Emotet & HarHar & Lokibot & njRAT & Qakbot & Swisyn & Trickbot & Ursnif & Zeus \\\hline
        $704$ & $14,429$ & $655$ & $4,191$ & $3,372$ & $4,895$ & $12,591$ & $4,202$ & $1,343$ & $2,594$ \\\hline
    \end{tabular}
    \caption{Number of samples in the dataset according to the malware family.}
    \label{tab:my_label}
\end{table}

\begin{figure}[t]
    \centering
    \includegraphics[width=1.0\textwidth]{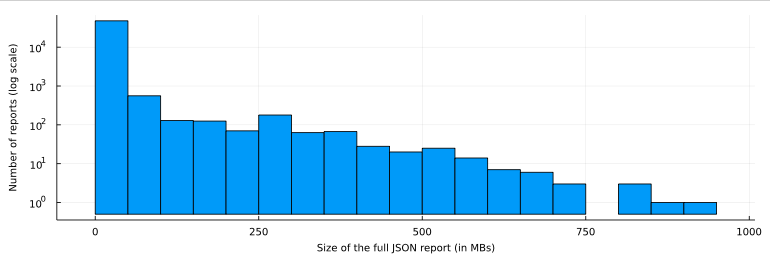}
    \caption{The histogram of sizes (in MBs) of full JSON reports produced by CAPEv2 on samples in our dataset.}
    \label{fig:sizes}
\end{figure}

Moreover, the samples are one of 6 types: 
\begin{center}
\begin{verbatim}
        "banker", "trojan", "pws", "coinminer", "rat", "keylogger"
\end{verbatim}
\end{center}
The samples were collected mainly throughout the years 2017 and 2019 (see Figure~\ref{fig:times} for the histogram of years and months in which the samples were detected) thus offering a significant spread over time for studying changes in families and the changes in data distribution. 

\subsection{Dataset}
The dataset itself consists of JSON reports produced by the CAPE sandbox. For each analyzed sample, there is a full JSON report of the CAPE sandbox, stored as a plain JSON in a file identified by the \texttt{sha256} of the analyzed sample. The full JSON files contain all gathered information, including all information about the started processes and system calls, all the generated (dumped) binaries, and also (if any) existing detections based on YARA language~\cite{cohen2017scanning}. 

Although the key purpose for releasing the dataset is to allow AI/ML researchers to develop and test new methods and classifiers, we are not providing any transformation of JSON reports into any vector representation. The main reason is that there exist methods that can take the JSON as an input without any transformation (e.g., using hierarchical multi-instance learning (HMIL) paradigm~\cite{HMIL} or treating JSONs as text documents~\cite{neurlux}). We follow the first approach and use HMIL models to train a classifier that takes JSONs as input.

However, the presence of some keys in the full reports can leak true labels and (e.g., detection based on YARA) thus invalidate using full reports directly for training ML models. Moreover, the size of the full reports can be prohibitively large -- in a few cases, the size of the full report is more than 800 MBs (see Figure~\ref{fig:sizes} for a full histogram of sizes of full reports). 

Therefore, besides the set of full reports, we have also created a set of reduced reports. We generate these reduced reports by restricting the original JSON to two keys containing enough information for training an ML classifier. These keys are:
\begin{itemize}
    \item \texttt{behavior $\rightarrow$ summary} that contains a summary of the behavioral analysis, including accessed files, registry keys, mutexes, and API calls, 
    \item \texttt{static $\rightarrow$ pe} that contains static properties of the executed sample.
\end{itemize}
An example of such a reduced report can be seen in Figure~\ref{fig:reduced_JSON_example}. In the following section, we introduce two baseline HMIL models that take these reduced JSON files as input and their task is to predict the correct malware family. 

\newpage 
\thispagestyle{empty}
\begin{figure}[H]
\vspace{-1.5cm}
    \centering
    \begin{lstlisting}
{
   "behavior": {
        "summary": {
            "keys": [
                "HKEY\_LOCAL\_MACHINE\\System\\CurrentControlSet\\Control\\Nls\\CustomLocale",
                ...], 
            "resolved\_apis": [
                "kernel32.dll.GetCurrentProcessorNumber",
                "kernel32.dll.GetNativeSystemInfo",
                ...], 
            "executed\_commands": [
                "\"C:\\Users\\comp\\AppData\\Local\\Temp\\FFFF450D574E5E5706FB.exe\"",
                "\"C:\\Users\\comp\\AppData\\Local\\Microsoft\\Windows\\spcmachine.exe\""], 
            "write\_keys": [],
            "files": [
                "C:\\Windows\\SysWOW64\\kernel32.dll",
                "C:\\Windows\\Globalization\\Sort,
                ...],
            ...
            "mutexes": [
                "PEMB40",
                "PEM868",
                "Global\\I5C3A8244",
                "Global\\M5C3A8244",
                ...]
        }
    },
    "static": {
        "pe": {
            "icon\_hash": null,
            "sections": [
                {
                    "raw\_address": "0x00001000",
                    "name": ".text",
                    "characteristics": "IMAGE\_SCN\_CNT\_CODE|IMAGE\_SCN\_MEM\_EXECUTE|IMAGE\_SCN\_MEM\_READ",
                    "virtual\_size": "0x00002786",
                    "virtual\_address": "0x00001000",
                    "size\_of\_data": "0x00003000",
                    "entropy": "5.83",
                    "characteristics\_raw": "0x60000021"
                },
            ...
            "peid\_signatures": null,
            "entrypoint": "0x00403600",
            "exports": [],
            "overlay": null,
            "digital\_signers": [],
            "imphash": "4e77bf5b96ea24734ed70b788b9fb7c8",
            "reported\_checksum": "0x00000000",
            "icon": null,
            "guest\_signers": {
                "aux\_error": true,
                "aux\_sha1": null,
                "aux\_timestamp": null,
                "aux\_valid": false,
                "aux\_signers": [],
                "aux\_error\_desc": "No signature found. SignTool Error File not valid 
                    C\\Users\\comp\\AppData\\Local\\Temp\\FFFF450D574E5E5706FB.exe"
            },
            "actual\_checksum": "0x0002a738",
            "imports": [
                {
                    "dll": "USER32.dll",
                    "imports": [
                        {
                            "name": "GetFocus",
                            "address": "0x404024"
                        }
                        ...],
			        "versioninfo": [],
                    "resources": [
                        {
                            "name": "RT\_STRING",
                            "filetype": null,
                            "offset": "0x000210a0",
                            "language": "LANG\_NORWEGIAN",
                            "sublanguage": "SUBLANG\_NORWEGIAN\_BOKMAL",
                            "size": "0x00000024",
                            "entropy": "0.55"
                        }],
                    "pdbpath": null,
                    "osversion": "6.0",
                    "icon\_fuzzy": null,
                    "imagebase": "0x00400000",
                    "imported\_dll\_count": 4,
                    "timestamp": "1995-11-19 14:43:13"
                }
        }}
}

    \end{lstlisting}
    \caption{Shortened example of a JSON output of CAPEv2 analysis of a malicious sample.}
    \label{fig:reduced_JSON_example}
\end{figure}

\newpage

\section{Baseline Performance of a Machine Learning Classifier}
As indicated above, we are using hierarchical multi-instance learning (HMIL) models~\cite{HMIL} to train classifiers that given a reduced JSON input predict the malware family of this sample. 

To train and test the accuracy of the baseline model, we split the dataset into a train and test part.
Following the practice of other malware-detection datasets~\cite{anderson2018ember}, we suggest researchers make the train-test split according to some date -- in our baseline case, all samples before $2019/08/01$ are considered to be a part of the training set ($37,512$ samples ($\approx76\%$)), all samples that appeared later are considered to be a part of the test set ($11,464$ samples ($\approx24\%$)). 
As we shall see, such a train-test split according to time is important as it takes into consideration the significant concept drift present in malware data.

We have created two HMIL models:
\begin{itemize}
    \item a \textbf{reduced model} that considers complete reduced JSON reports as described in the previous section
    \item a \textbf{static reduced model} that considers only the static part of the reduced JSON reports.
\end{itemize}
For training the HMIL models, we have used a public library written in Julia Mill.jl~\cite{mandlik2021milljl} together with JsonGrinder.jl~\cite{mandlik2021milljl} (the jupyter notebooks demonstrating the training of the baseline models are part of the released dataset).
We have used the default setting while creating the model for the reports (i.e., the size of the inner dense layers (set to 32), the aggregation function (set to \texttt{meanmax} aggregation function), and the activation function (set to \texttt{relu}).
Using $200$ training steps, each with randomly sampled minibatch of size $500$ and \texttt{ADAM} optimizer, the trained reduced model achieved an accuracy of $99.5\%$ on training data and $94.5\%$ on testing data. The complete results, i.e., the accuracy per different malware families, are shown in Table~\ref{tab:CM_complete}.

\begin{table}[h]
\footnotesize
    \centering
    \begin{tabular}{|c||r|r|r|r|r|r|r|r|r|r|}
    \hline
    True \textbackslash ~Pred.	&   Adload&   Emotet&   HarHar&  Lokibot&   Qakbot&   Swisyn& Trickbot&   Ursnif&     Zeus&    njRAT \\\hline\hline
Adload	 & \textbf{100.00} &      0.00 &      0.00 &      0.00 &      0.00 &      0.00 &      0.00 &      0.00 &      0.00 &      0.00\\\hline
  Emotet	 &      0.06 & \textbf{86.54} &      0.00 &      0.48 &      0.00 &      0.03 &      3.09 &      2.30 &      7.28 &      0.23\\\hline
  HarHar	 &      0.00 &      0.00 & \textbf{98.55} &      0.00 &      0.00 &      0.00 &      0.00 &      0.00 &      0.00 &      1.45\\\hline
 Lokibot	 &      0.00 &      0.15 &      0.00 & \textbf{95.33} &      0.00 &      0.00 &      0.15 &      0.00 &      0.15 &      4.23\\\hline
  Qakbot	 &      0.00 &      0.00 &      0.00 &      0.25 & \textbf{99.24} &      0.00 &      0.00 &      0.00 &      0.51 &      0.00\\\hline
  Swisyn	 &      0.00 &      0.00 &      0.00 &      0.00 &      0.00 & \textbf{99.88} &      0.00 &      0.02 &      0.02 &      0.07\\\hline
Trickbot	 &      0.00 &      0.15 &      0.00 &      0.22 &      0.00 &      0.00 & \textbf{98.58} &      0.15 &      0.52 &      0.37\\\hline
  Ursnif	 &      0.00 &      0.00 &      0.00 &      0.20 &      0.20 &      0.00 &      0.00 & \textbf{94.02} &      4.18 &      1.39\\\hline
    Zeus	 &      0.00 &      1.39 &      0.00 &      0.00 &      0.00 &      0.00 &      0.00 &      0.00 & \textbf{95.83} &      2.78\\\hline
   njRAT	 &      0.00 &      0.00 &      0.00 &      0.18 &      0.00 &      0.00 &      0.00 &      0.18 &      0.37 & \textbf{99.26}\\\hline
   \end{tabular}
    \caption{The confusion matrix of predictions on testing data for the HMIL model trained on reduced JSON reports. The ground-truth labels are provided in rows, the predictions are provided in columns, and every cell represents a percentage of samples having the true label (the row label) being classified as the label in the column.}
    \label{tab:CM_complete}
\end{table}

To confirm the importance of behavioral data, we have trained also a \emph{static} model that considers only the static part of the reduced JSON reports. Using the same parameters (i.e., the parameters for the model and the training process), the trained model reached an accuracy of $96.7\%$ on training data. However, the model does not generalize in the face of the concept drift and the accuracy of the model on the testing data was (at best) around $63\%$. From the details in Table~\ref{tab:CM_static} it is clear that mainly malware families Emotet and Qakbot were exhibiting a significant change in static characteristics such that the model was not able to detect it on the testing set (only $2.35\%$ of correctly predicted Emotet samples and $35.63\%$ of Qakbot samples, respectively). Finally, note that without the time split, i.e., if all data were randomly split into training and testing sets regardless of the time stamp, the average accuracy of the static reduced model was well above 95\%. This demonstrates the necessity of considering a time split since otherwise, the model will leverage easily identifiable features that, however, do not generalize over time and thus are not really solving the malware-detection problem.

\begin{table}[t]
\footnotesize
    \centering
    \begin{tabular}{|c||r|r|r|r|r|r|r|r|r|r|}
    \hline
True \textbackslash ~Pred.	 &   Adload &   Emotet &   HarHar &  Lokibot &   Qakbot &   Swisyn & Trickbot &   Ursnif &     Zeus &    njRAT\\\hline\hline
  Adload	 & \textbf{60.00} &      0.00 &      0.00 &      0.00 &      0.00 &      0.00 &     20.00 &     20.00 &      0.00 &      0.00\\\hline
  Emotet	 &      0.00 & \textbf{2.35} &      0.00 &     29.13 &      1.70 &      0.00 &     57.30 &      6.46 &      1.79 &      1.28\\\hline
  HarHar	 &      0.00 &      0.00 & \textbf{98.55} &      0.00 &      0.00 &      0.00 &      0.00 &      0.00 &      0.00 &      1.45\\\hline
 Lokibot	 &      0.00 &      0.15 &      0.00 & \textbf{90.07} &      0.29 &      0.00 &      0.73 &      0.29 &      0.58 &      7.88\\\hline
  Qakbot	 &      0.00 &     54.45 &      0.00 &      4.33 & \textbf{35.62} &      0.00 &      0.51 &      4.07 &      1.02 &      0.00\\\hline
  Swisyn	 &      0.02 &      0.00 &      0.00 &      0.05 &      0.00 & \textbf{99.72} &      0.14 &      0.02 &      0.02 &      0.02\\\hline
Trickbot	 &      0.00 &      0.90 &      0.00 &     16.47 &      0.30 &      0.00 & \textbf{68.49} &      4.49 &      0.60 &      8.76\\\hline
  Ursnif	 &      1.79 &     20.92 &      0.00 &      1.59 &      0.20 &      0.00 &      1.59 & \textbf{70.32} &      2.99 &      0.60\\\hline
    Zeus	 &      0.00 &      5.56 &      0.00 &      2.78 &      0.00 &      0.00 &      0.00 &      7.64 & \textbf{82.64} &      1.39\\\hline
   njRAT	 &      0.00 &      0.74 &      0.00 &     13.44 &      0.00 &      0.00 &      0.74 &      4.60 &      4.97 & \textbf{75.51}\\\hline
   
   \end{tabular}
    \caption{The confusion matrix of predictions on testing data for the \textbf{static} HMIL model trained on reduced JSON reports. The ground-truth labels are provided in rows, the predictions are provided in columns, and every cell represents a percentage of samples having the true label (the row label) being classified as the label in the column.}
    \label{tab:CM_static}
\end{table}

\section{Discussion and Conclusions}
The released dataset presents a unique collection of detailed behavioral data collected from malicious samples over an extensive period of time. We have shown that the released data contain enough information so that a reasonably accurate classifier can be trained, however, we believe that the released dataset offers a variety of new research opportunities. 

Given the long period over which the samples were collected, the dataset should be useful for studying concept drift and changes in malicious behavior. Connected to this, a challenge is to train a classifier to identify the malicious behavior (or attribution to a specific malware family) in a way robust to concept drift by identifying generic techniques and tactics that may be hidden in the full list of system calls of complete behavioral logs.

\bibliographystyle{abbrv}
\bibliography{refs.bib}

\end{document}